\documentclass[12pt,reqno]{amsart}

\usepackage{amssymb,amsxtra}
\usepackage{amsfonts}
\usepackage{amsmath} 
\usepackage{graphicx} 
\usepackage{layout}

\newcommand{\erre}{\mathbb{R}} 

\newcommand{\natu}{\mathbb{N}}

\newcommand{\ve}{\varepsilon}

\newcommand{\ih}{\frac{i}{\hbar}}

\newcommand{\n}{\noindent} 
\newcommand{\vs}{\vspace{0.5cm}}
\newcommand{\f}{\frac}
\newcommand{\ba}{\begin{eqnarray}} 
\newcommand{\ea}{\end{eqnarray}}
\newcommand{\be}{\begin{equation}}
 \newcommand{\ee}{\end{equation}}
\newcommand{\bdm}{\begin{displaymath}}
\newcommand{\edm}{\end{displaymath}} 
\newcommand{\brr}{\begin{array}}
\newcommand{\err}{\end{array}}

\newcommand{\bml}{\begin{gather}} \newcommand{\eml}{\end{gather}}

\newcommand{\spaz}{\vspace{.5cm} \noindent}

\numberwithin{equation}{section}

\begin{document}

\title[ ]{
Joint excitation probability for two harmonic \\
\vspace{0.1cm}oscillators in dimension one and the Mott problem\\ \vspace*{0.6cm}}

\author{Gianfausto dell'Antonio}
\address{Dell'Antonio: Dipartimento di Matematica, Universit\'a di Roma "La Sapienza"}

\curraddr{P.le A. Moro, 2 - 00185 Roma, Italy}

\email{dellantonio@mat.uniroma1.it}

\author{Rodolfo Figari} 
\address{ Figari: Dipartimento di Scienze Fisiche,
Universit\'a di Napoli and Sezione I.N.F.N. Napoli}

\curraddr{Via Cinthia 45, 80126 Napoli, Italy}

\email{figari@na.infn.it}

\author{Alessandro Teta} \address{ Teta: Dipartimento di Matematica Pura ed
Applicata, Universit\`a di L'Aquila}

\curraddr{Via Vetoio (Coppito 1), 67010 L'Aquila, Italy}

\email{teta@univaq.it}

{\maketitle}

\vs

\begin{abstract}
We analyze a one dimensional quantum system consisting of a test particle interacting with two harmonic oscillators placed at the positions $a_1$, $a_2$, with $a_1 >0$, $|a_2|>a_1$, in the two possible situations: $a_2>0$ and $a_2 <0$.  At time zero the harmonic oscillators are in their ground state and the test particle is in a superposition state of two wave packets centered in the origin with opposite mean momentum. 
Under suitable assumptions on  the physical parameters of the model, 
we consider the time evolution of the wave function and we compute 
the probability $\mathcal{P}^{-}_{n_1 n_2} (t)$ (resp. $\mathcal{P}^{+}_{n_1 n_2} (t)$)  that both oscillators are in the excited states labelled by $n_1$, $n_2 >0$ at time $t >  |a_2| v_0^{-1}$ when $a_2 <0$ (resp. $a_2 >0$).

\n
We prove   that $\mathcal{P}_{n_1 n_2}^- (t)$ is negligible with  respect to  $\mathcal{P}_{n_1 n_2}^+ (t)$, up to second order in time dependent perturbation theory.

\n
The system we consider is   a simplified, one dimensional version of the original model of a cloud chamber introduced by Mott in  \cite{m},  where the result was argued using euristic arguments in the framework of the time independent perturbation theory for the stationary Schr\"{o}dinger equation.

\n

\n

\n
The method of the proof is entirely elementary and it is essentially based on a stationary phase argument. We also remark that all the computations refer to the Schr\"{o}dinger equation for the three-particle system, with  no reference to  the  wave packet collapse postulate.

\end{abstract}

\vs

\section{Introduction, notation and result }

\vs
\n
In his paper of 1929 Mott  (\cite{m}) analyzes the dynamics of formation  of tracks left  an $\alpha$-particle emitted by a radioactive source inside the supersaturated  vapour in a cloud chamber. He notices
the  difficulty to understand intuitively 
how a spherical wave function,
describing the particle isotropically emitted by the source, might
manifest itself as a straight track in the cloud chamber.

\n
Without referring to any wave packet collapse, he proposes an explanation
based on the analysis of the 
whole quantum system made up
of the $\alpha$-particle and of the atoms of the vapour. Using a simplified model with only two atoms 
and making use of time independent perturbation arguments, he concludes that
each ionization process focuses  the probability of
presence of the $\alpha$-particle 
on narrower and narrower cones, around the
straight line connecting the source to the ionized atoms.

\n
In this way Mott suggests a quantum dynamical mechanism responsible of the
transition between an initial superposition of outgoing waves heading isotropically in all directions 
toward an incoherent (classical) sum of those same waves.

\n
We mention that the same problem is also discussed in \cite{h} and later in \cite{be}, where the above approach is compared with the   explanation based on the wave packet collapse. We refer to \cite{b}, \cite{ha}, \cite{cl}, \cite{p}  for some further elaborations  on the subject and to \cite{lr} for  a  description of the original experimental apparatus.

\n
The aim of our work is to
provide a detailed time dependent analysis of a one dimensional version of the system
investigated by Mott.  The system we consider consists of  
 a test particle 
and two harmonic oscillators. In our model a superposition of two wave packets centered in the origin with opposite momentum  plays the role of the spherical wave of the $\alpha$-particle and the oscillators replace the atoms to be ionized. Under suitable assumptions on the physical parameters of the model,  we   perform  a detailed time analysis of the
evolution of the system wave function using time dependent perturbation
theory and we give a quantitative estimate of the joint excitation probability of the oscillators. Roughly speaking, our main result is that such probability is essentially zero if the oscillators are placed on opposite sides of  the origin, while it has a finite,  non-zero value in the other case. Following the line of reasoning of Mott, the result can be interpreted saying that before the interaction the test particle is delocalized while after the interaction it is either on the left (if there is an excited oscillator on the left) or on the right (if there   is an excited oscillator on the right). In any case one can say that the test particle propagates along an almost classical trajectory, without making any reference to the wave packet collapse postulate.

\n
In \cite{ccf} the authors consider a similar problem in three dimensions where a
particle interacts via zero range forces with localized two level quantum
systems. A non perturbative  analysis of the model is carried out but
results are valid only in the scattering regime.

\n
Let us introduce the model. We consider a three-particle non relativistic quantum system in dimension one, made of  one test particle   with mass $M$ interacting with  two harmonic oscillators  with the identical mass $m$. 
We denote by $R$ the position coordinate of the test particle and by $r_1$,$r_2$ the position coordinates of the two oscillators. 
The Hamiltonian of the system  in $L^2(\erre^3)$ is
\ba
&&H = H_0 + \lambda H_1 \label{H}\\
&&\nonumber\\
&&H_0= -\f{\hbar^2}{2M} \Delta_R -\f{\hbar^2}{2m} \Delta_{r_1}  +\f{1}{2}m \omega^2 (r_1 - a_1)^2 - \f{\hbar^2}{2m} \Delta_{r_2} +\f{1}{2}m \omega^2 (r_2 - a_2)^2 \label{H_0}\\
&&\nonumber\\
&&H_1 = V(\delta^{-1}(R-r_1)) + V(\delta^{-1}(R-r_2)) \label{H_1}
\ea
where $\lambda >0$, $\omega >0$, $a_1 >0$, $a_2 \in \erre$,  with $a_1 < |a_2|$, $\delta >0$ and $V$ is a  smooth interaction potential. 
The assumptions on $V$ will guarantee that  the Hamiltonian  $H$ is self-adjoint with the same  domain of $H_0$ and then the evolution problem corresponding to the Hamiltonian $H$ is well posed. 
 For the test particle we choose an initial state $\psi$ in the form of a superposition state
\ba
&&\psi(R)= \psi^{+}(R) + \psi^{-}(R)\label{statoin}\\
&&\psi^{\pm}(R )= \f{\mathcal{N}}{\sqrt{\sigma}} e^{- \f{R^2}{2 \sigma^2}} e^{\pm i \f{P_0}{\hbar} R}, \;\;\;\;\;\;P_0 = M v_0
\label{statoin2}
\ea
where $\sigma >0$, 
${\mathcal N}= \big[ 2 \sqrt{\pi}  (1 + e^{- \left(\f{P_0 \sigma}{\hbar} \right)^2 } )\big]^{-1/2}$
is the normalization factor and $P_0, v_0$ denote the absolute value of the initial mean momentum and velocity of the test particle.

\n
For the harmonic oscillator centered in $a_j$, $j=1,2$, the initial state $\phi_0^{a_j}$ is the corresponding ground state. Moreover we define 
\ba\label{aosc}
&&\phi^{a_j}_{n_j}(r_j)= \f{1}{\sqrt{\gamma}} \phi_{n_j} (\gamma^{-1}(r_j -a_j))\\
&&\gamma = \sqrt{\f{\hbar}{m \omega}}\label{gamma}
\ea
where 
$\phi_m$ is the normalized Hermite function of order $m \in \natu$. We  notice that the parameter $\gamma$  has the dimension of a length and  gives  a measure of the spatial localization of the oscillators.

\n
Let us denote by $\Psi(R,r_1,r_2,t)$ the wave function of the system;  $\Psi(t)\equiv \Psi(\cdot,\cdot,\cdot,t)$ is the solution of the  Cauchy problem
\ba
&&i \hbar \f{\partial }{\partial t} \Psi(t) = H \Psi(t) \label{eq}\\
&&\nonumber\\
&&  \Psi(0)= \psi \, \phi^{a_1}_{0} \, \phi^{a_2}_{0} \label{statoin}
\ea

\n
We are interested in the probability that both harmonic oscillators are in an excited state at a given time $t>0$.  The solution of the three-body problem (\ref{eq}),(\ref{statoin})  is not known in closed form;   we shall limit ourselves to a perturbative computation. It is worth mentioning   that,   in order to get a non trivial result, we are forced to compute the second order approximation of the solution of the Cauchy problem 
(\ref{eq}),(\ref{statoin}), which we denote by $\Psi_2(R,r_1,r_2,t)$.   Therefore the object of our analysis is the quantity

\ba\label{prob}
&&\mathcal{P}^{\pm}_{n_1 n_2 }(t) = \int \! dR \left| \int \! dr_1 dr_2 \, \phi^{a_1}_{n_1} (r_1) 
\phi^{a_2}_{n_2} (r_2) \Psi_2(R,r_1,r_2,t) \right|^2
\ea 
for $n_1 \neq 0$ and $n_2 \neq 0$, where $\pm$ refers to the cases $a_2 >0$ and $a_2 <0$ respectively. Formula (\ref{prob}) represents the probability that both oscillators are in an excited state at time $t$, up to second order in perturbation theory.

\n
The explicit computation of (\ref{prob}) will be performed exploiting some 
 further assumptions on the physical parameters of the model.  More precisely the complete set of assumptions required for our analysis is the following

 \vs
\n
\hspace*{1cm}$(A_0)$ 
\ba
&&\lambda_0 \equiv \f{\lambda}{Mv_0^2} \ll 1
\ea
\vs
\n
\hspace*{1cm}$(A_1)$ 

\n
{\em The  quantities  \hspace{0.3cm}$\delta m \equiv  \f{m}{M}, \;\;\; \delta E \equiv \f{\hbar \omega}{M v_0^2},\;\;\;\delta R \equiv \f{\sigma}{|a_j|},\;\;\;
\delta L \equiv \f{\delta}{|a_j|} , \;\;\; \delta \tau_j \equiv \f{v_0}{\omega |a_j|}$, for $j=1,2$, are all  $O(\ve)$ where}
\be\label{la0}
\lambda_0 \ll \ve \ll 1
\ee

\vs
\hspace*{0.2cm}$(A_2)$

\vspace{0.3cm}
\n
{\em The interaction potential $V : \erre \rightarrow \erre$ is a continuous, positive and compactly supported  function.}

\vs
\n
Let us briefly comment on the above assumptions. In  $(A_0)$ we 
ensure  that the dimensionless coupling constant $\lambda_0$ is small.    In $(A_1)$ we assume that  the mass and  the kinetic energy of the test particle are much larger than the mass and the spacing of the energy levels of the oscillators; moreover  the initial wave packets of the test particle are assumed  to be well localized and the interaction is required to be short range; finally  the characteristic time of the oscillators $\omega^{-1}$ is assumed to be much smaller than  the flight times $\tau_1$, $\tau_2$ of the test particle, which  are defined by 
 \ba\label{tau12}
 &&\tau_1 = \f{a_1}{v_0}, \;\;\;\;\;\; \tau_2=\f{|a_2|}{v_0}
 \ea
Condition (\ref{la0}) guarantees that the first and second order corrections  in perturbation theory  remain  small compared with the unperturbed wave function, in fact of order $\lambda_0 \ve^{-1}$ and $\lambda_0^2 \ve^{-2}$ respectively.

\n
In order to understand the meaning of  $(A_1)$, let us consider the parameters   $M$, $v_0$, $a_1$ $a_2$ all of order one.  Then we obtain $m =O(\ve)$, $\omega=O(\ve^{-1})$, $\hbar \omega = O(\ve)$, $\sigma=O(\ve)$, $\delta=O(\ve)$.

\n
We  observe that the length $\gamma$ introduced in (\ref{gamma}) can be written as 
\be\label{gamma1}
\gamma=a_1 \,\delta \tau_1 \sqrt{\f{\delta E}{\delta m}}
\ee
and this means that $\gamma$ is of the same order of $\delta$ and $\sigma$. In particular this guarantees that the transit time of the test particle on the region where the oscillators  are localized is of the same order of the characteristic time of the oscillators.  To simplify the notation, from now on we shall fix
 \be\label{d=g}
 \delta = \gamma
 \ee

\n
We also introduce here a (large) parameter which is useful to express the various estimates in the proof

\be\label{lamj}
\Lambda_j \equiv \f{|a_j|}{\gamma}= O(\ve^{-1}) , \;\;\;\; j=1,2
\ee

\n
Our main result is the following.

\vs
\n
{\bf Theorem 1.} {\em Let us assume $(A_0)$, $(A_1)$, $(A_2)$ and fix $t>\tau_2$, $n_1 \neq 0$, $n_2 \neq 0$. Then for any $k \in \natu$, with $k>2$, we have 

\ba\label{p-}
&&\mathcal{P}^{-}_{n_1 n_2} (t) \leq \, \f{1}{\Lambda_1^{2k -4}}\left(\! \f{\lambda_0}{
\varepsilon} \! \right)^{\!\!4}  C_{n_1 n_2}^{(k)} (t)\\
&&\label{p+}
\mathcal{P}^{+}_{n_1 n_2} (t)=16 \pi^4 \sqrt{\pi} \left(\!\f{\lambda_0}{
\varepsilon} \! \right)^{\!\!4} \mathcal{N}^2 \left| \prod_{j=1,2} \tilde{V}\!(q_j)  (\widetilde{\phi_{n_j} \phi_0}) (q_j)\right|^{2} + \mathcal{S}_{n_1 n_2} (t)\\
&&q_j = - n_j \sqrt{\f{\delta E}{\delta m}}\label{qj} \\
&&|\mathcal{S}_{n_1 n_2} (t)| \leq \f{1}{\Lambda_1} \left(\! \f{\lambda_0}{\varepsilon
} \! \right)^{\!\!4} D_{n_1 n_2} (t)\label{p++}
\ea
where   
the symbol \hspace{0.02cm} $\widetilde{}$ \hspace{0.01cm}  
denotes  Fourier transform  and  $ C_{n_1 n_2}^{(k)} (t)$, $D_{n_1 n_2} (t)$ are   functions  of the physical parameters of the model which will be explicitely given during the proof (see (\ref{P-3}), (\ref{Dt}) below).

}

\vs

\n
We remark that the estimates (\ref{p-}), (\ref{p+}), (\ref{p++}) are not optimal; in particular $ C_{n_1 n_2}^{(k)} (t)$ and  $D_{n_1 n_2} (t)$ diverge for $t \rightarrow \infty$. From (\ref{P-3}), (\ref{Dt}) it will be clear that $C^{(k)}_{n_1 n_2}(t)$, $D_{n_1 n_2} (t)$ are of order one, and then  the estimates   are meaningful  only for $t$ larger but  of the same order of $\tau_2$.

\n
Let us briefly outline the strategy of the proof  and give a   heuristic argument which, at least at a qualitative level, justifies the result stated in theorem 1.     We find  convenient to represent the solution of (\ref{eq}), (\ref{statoin}) in the form 

\ba\label{exp}
&&\Psi(R,r_1,r_2,t) = \sum_{n_1,n_2} f_{n_1 n_2}(R, t) \phi^{a_1}_{n_1}(r_1) \phi^{a_2}_{n_2}(r_2)
\ea

\n
where    $f_{n_1 n_2}(\cdot ,t) =f_{n_1 n_2}(t)$ belongs to $L^2(\erre)$ for any $n_1,n_2 \in \natu$ and $t \geq 0$, and it is explicitely given by
\ba\label{fou}
&&f_{n_1 n_2}(R,t) = \int \! dr_1 dr_2  \phi^{a_1}_{n_1}(r_1) \phi^{a_2}_{n_2}(r_2)\Psi(R,r_1,r_2,t)
\ea

\n


\n
We notice that the coefficients of the expansion $f_{n_1n_2} (R,t) $ have a precise physical meaning;  
in fact the quantity
\be\label{expr}
\int_{\Omega} \!dR \, |f_{n_1n_2} (R,t) |^2
\ee

\n
represents the probability to find at time $t$ the test particle in $\Omega \subseteq \erre$ when the two oscillators are in  the states labelled by $n_1$,$n_2$ respectively.

\n
The equation for the coefficients $f_{n_1 n_2}(t)$ is obtained from Duhamel's formula

\ba\label{du}
&&\Psi(t)= e^{-\ih t H_0}\Psi_0
-i \f{\lambda}{\hbar}\int_{0}^{t} \! ds \,  e^{- \ih (t-s)H_0} H_1 
\Psi(s)
\ea

\n
 multiplying by $\phi^{a_1}_{n_1} \phi^{a_2}_{n_2}$ and then integrating with respect to the coordinates of the oscillators. The result is

\ba\label{duf}
&& f_{n_1 n_2}(t)=f^{(0)}_{n_1 n_2}(t) - \int_{0}^{t} \! ds \, \Gamma_{n_1 n_2}(t-s) \left( \sum_{j_1} V^{a_1}_{n_1 j_1} f_{j_1 n_2}(s) + \sum_{j_2} V^{a_2}_{n_2 j_2} f_{n_1 j_2}(s) \right)
\ea

\n
where in the above formula we have introduced the notation

\ba
&&f^{(0)}_{n_1 n_2} (t) = \delta_{n_1 0}\, \delta_{n_2 0} \, e^{-2 \ih t E_0} \, e^{- \ih t K_0} \psi\label{f0}\\
&&K_0 = - \frac{\hbar^2}{2M} \Delta_{R}\\
&&\Gamma_{n_1 n_2} (t) = i \f{\lambda}{\hbar} e^{- \ih t(E_{n_1} + E_{n_2})} \, e^{- \ih t K_0}\\
&&V_{m n}^{a_i} (x)= \int \! dy \, \phi_{m}^{a_i}(y) \phi_{n}^{a_i}(y) V(\gamma^{-1}(x-y)), \;\;\;\; m,n \in \natu, \;\; i=1,2
\ea

\n
We want to give an estimate  of the solution of (\ref{duf}) up to second order in perturbation theory. Iterating twice equation (\ref{duf}) we obtain

\ba\label{f012}
&&f_{n_1 n_2} (t)=f^{(0)}_{n_1 n_2} (t)+ f^{(1)}_{n_1 n_2} (t) + f^{(2)}_{n_1 n_2} (t)+ \mathcal{E}_{n_1 n_2}(t)
\ea

\n 
where 
\ba\label{f1}
&&f^{(1)}_{n_1 n_2}(t) = - \int_{0}^{t} \! ds \, \Gamma_{n_1 n_2}(t-s) \left( \sum_{j_1} V^{a_1}_{n_1 j_1} f^{(0)}_{j_1 n_2}(s) + \sum_{j_2} V^{a_2}_{n_2 j_2} f^{(0)}_{n_1 j_2}(s) \right)\\
&&f^{(2)}_{n_1 n_2}(t) = - \int_{0}^{t} \! ds \, \Gamma_{n_1 n_2}(t-s) \left( \sum_{j_1} V^{a_1}_{n_1 j_1} f^{(1)}_{j_1 n_2}(s) + \sum_{j_2} V^{a_2}_{n_2 j_2} f^{(1)}_{n_1 j_2}(s) \right) 
\label{f2}
\ea

\n
and $\mathcal{E}_{n_1 n_2}(t)$ is the error term which we shall neglect in the sequel. Obviously we have 

\be
\Psi_2 (R,r_1,r_2, t) = \sum_{n_1 ,n_2} \left(  f^{(0)}_{n_1 n_2} (R,t)+ f^{(1)}_{n_1 n_2} (R,t) + f^{(2)}_{n_1 n_2} (R,t) \right) \phi^{a_1}_{n_1}(r_1) \phi^{a_2}_{n_2}(r_2)
\ee

\n
Exploiting the explicit expression (\ref{f0}) of $f^{(0)}_{n_1 n_2}(t)$, we can write $f^{(1)}_{n_1 n_2}(t)$ in the form

 \ba\label{2f1}
&&f^{(1)}_{n_1 n_2}(t)\nonumber\\
&&= - \delta_{n_2 0} \! \int_{0}^{t} \!\!\!ds \, \Gamma_{n_1 0}(t-s)\, V^{a_1}_{n_1 0}\, \,  e^{- 2 \ih s E_0} \, e^{- \ih s K_0} \psi 
- \delta_{n_1 0}\! \int_{0}^{t} \!\!\!ds \, \Gamma_{0 n_2 }(t-s)\, V^{a_2}_{n_2 0}\,\,  e^{- 2 \ih s E_0} \, e^{- \ih s K_0} \psi  \nonumber\\
&&\equiv \delta_{n_2 0}\, f^{(1)}_{n_1 0}(t) + \delta_{n_1 0}\, f^{(1)}_{ n_2 0}(t)
\ea

\vs

\n
From formula (\ref{2f1}) it is clear that $f^{(1)}_{n_1 n_2}(t)=0$  if $n_1 \neq 0$ and $n_2 \neq 0$. As expected, this means that the probability that both oscillators are in an excited state is zero up to first order in perturbation theory. As a consequence, from (\ref{prob}) we get

\be\label{prob2}
\mathcal{P}^{\pm}_{n_1 n_2} (t) = \int \! dR \left| f^{(2)}_{n_1 n_2} (R,t) \right|^2 , \;\;\;\;\;n_1 \neq 0, \; n_2 \neq 0
\ee

\n
\n
Following  the original strategy of Mott, a crucial point of the analysis is the explicit  evaluation of  $f^{(1)}_{n_1 0}(t)$ and $f^{(1)}_{ n_2 0}(t)$.  We notice that $V^{a_1}_{n_1 0} (x)$  and $\left(\! e^{-\f{i}{\hbar} s K_0} \psi^{\pm} \! \right)\!(x)$ are essentially different from zero only for $x \simeq a_1$ and $x \simeq \pm v_0 s$ respectively. This means that the only non zero contribution to the time integral defining $f^{(1)}_{n_1 0}(t)$ comes from $\psi^+$ and such contribution is essentially concentrated around $s = \f{a_1}{v_0}= \tau_1$. 
 Hence we can argue that  $f^{(1)}_{n_1 0}(t)$ is approximately given by a wave packet starting at time $\tau_1$ from the position $a_1$ of the first oscillator, with a velocity close to $v_0$. In particular it  is essentially different from zero only in a neighborhood of $a_1 +v_0 (t- \tau_1)$,   for $t > \tau_1$. 

\n
Analogously, $f^{(1)}_{ n_2 0}(t)$ is approximately given by a wave packet starting at time $\tau_2$ from the position $a_2$ of the second oscillator, with a velocity close to $v_0$ if $a_2 >0$, and to $-v_0$ if $a_2 <0$. Then $f^{(1)}_{ n_2 0}(t)$ is essentially different from zero only in a neighborhood of $a_2 +v_0( t- \tau_2)$,   for $t > \tau_2$,  $a_2 >0$, and  in a neighborhood of $a_2  - v_0 (t- \tau_2)$,   for $t > \tau_2$,  $a_2 <0$.

 \n
 Let us now consider  the second order term $f^{(2)}_{n_1 n_2}(t)$; 
exploiting expression (\ref{2f1}), we have

\ba\label{2f2}
&&f^{(2)}_{n_1 n_2}(t) = - \delta_{n_2 0} \! \int_{0}^{t} \! ds \, \Gamma_{n_1 0}(t-s)\sum_{j_1} V^{a_1}_{n_1 j_1} f^{(1)}_{j_1 0}(s)
 - \delta_{n_1 0}\!\int_{0}^{t} \! ds \, \Gamma_{0 n_2}(t-s)\sum_{j_2} V^{a_2}_{n_2 j_2} f^{(1)}_{0 j_2}(s)\nonumber\\
&&- \int_{0}^{t} \! ds \, \Gamma_{n_1 n_2}(t-s) V^{a_1}_{n_1 0}\, f^{(1)}_{0 n_2}(s)
- \int_{0}^{t} \! ds \, \Gamma_{n_1 n_2}(t-s) V^{a_2}_{n_2 0}\, f^{(1)}_{ n_1 0}(s)
\ea

\n
Since we are interested in the probability that both oscillators are excited, only the last two terms of (\ref{2f2}) are relevant.

\n
We notice that the supports of $V^{a_1}_{n_1 0}$ and $ f^{(1)}_{ n_2 0}(s) $ are essentially disjoint for any $s \geq 0$ and this implies that the third term in  the r.h.s. of (\ref{2f2}) gives a negligible contribution.

\n
For the same reason, the fourth term   in  the r.h.s. of (\ref{2f2}) is also approximately zero if $a_2 <0$. 

\n
This explains why we expect that an estimate like (\ref{p-}) holds.

\n
On the other hand, in the  case $a_2 >0$ the product  $V^{a_2}_{n_2 0}  f^{(1)}_{n_1 0}(s)$  is   different from zero for $s \simeq \tau_2$. In such case    the fourth term   in  the r.h.s. of (\ref{2f2}) gives a non zero contribution and this explains why we can expect that a formula like  (\ref{p+}) holds.

\n
We collect here some further notation which will be used in the paper.

\n
- $\langle x \rangle$ denotes $(1 + x^2)^{1/2}$;

\n
- $d_{x_l}^k f$ is the derivative of order $k$ with respect to $x_l$ of a smooth function  $f(x_1, \ldots ,x_n)$, for  

\n
\hspace*{0.15cm} $n \in \natu$ and $l=1, \ldots,n$;

\n
- $\|f\|_{W^{k,1}_{s}} =\sum_{l=1}^{n} \sum_{m=0}^{k} \int \! dx \, \langle x \rangle^s |(d_{x_l}^{m} f)(x)|$, $k \in \natu$, $s \geq 0$;

\n
- $\|f\|_{L^1_s}= \| f \|_{W^{0,1}_s}$;

\n
- $c$ is a generic positive numerical constant.

\n
The paper is organized as follows.  In section 2 we study the first order approximation step. In section 3 we analyze the second order approximation, distinguishing the two case $a_2 >0$ and $a_2 <0$. In section 4 we compute the joint excitation probability of the two oscillators concluding the proof of theorem 1. Finally in the appendix we give a proof of a technical lemma.


\section{First order approximation }

\vs

\n
In this section we fix $t> \tau_j$, $j=1,2$, and we give an estimate of the first order terms $f^{(1)}_{n_j 0}(t)$. We only give the details for the case $a_2 >0$ since the opposite case can be treated similarly. 
We rewrite  $f^{(1)}_{n_j 0}(t) $  as follows

\ba\label{fipm}
&& f^{(1)}_{n_j 0}(t) = f^{(1),+}_{n_j 0}(t) + f^{(1),-}_{n_j 0}(t)  \\
&&f^{(1),\pm}_{n_j 0}(t) = - \Gamma_{n_j 0}(t)  \int_{0}^{t} \! ds \;  e^{i n_j \omega s} \; e^{\f{i}{\hbar} s K_0} \; V_{n_j 0}^{a_j} \; e^{-\f{i}{\hbar} s K_0} \psi^{\pm} \label{fipm2}
\ea

\n
Moreover let us define for $j=1,2$ and $ s,t \geq 0$

\ba\label{Ij}
&&\mathcal{I}_j(s) = e^{\f{i}{\hbar} s K_0} \,V^{a_j}_{n_j 0} \;e^{- \f{i}{\hbar} s K_0} 
\\
&&h_{j}^{\pm}(t) =  \int_0^t \!\! ds \, e^{i n_j \omega s}\, \mathcal{I}_j (s) \psi^{\pm}\label{hj}
\ea

\vs
\n
 As a  first step  the operator (\ref{Ij})  will be written in a more convenient form.

\vs
\n
{\bf Lemma 2.1.} {\em For any $f \in L^2(\erre)$ and $s \geq 0$ the following identity holds

\ba\label{Ijf}
&&
(\mathcal{I}_j (s) f )(R)= \int \!\!  d \xi \; g_{j}(\xi) f (R+ (M \gamma)^{-1} \hbar s \, \xi)\, e^{ i \f{\hbar s}{2 M \gamma^2} \xi^2} \,    e^{i\f{R}{\gamma} \xi}  \; e^{-i \Lambda_j \xi} 
\ea
where}
\ba\label{gnj}
&&g_{j}(\xi)= \;\tilde{V} (  \xi) \;\widetilde{(\phi_{n_j} \phi_0)} (\xi) \;   
\ea

\vs
\n
{\bf Proof. } Exploiting the explicit expression of the free propagator we have

\ba\label{freep}
&& \left(  e^{\f{i}{\hbar} s K_0} \; V_{n_j 0}^{a_j} \; e^{-\f{i}{\hbar} s K_0} f  \right)\!(R) \nonumber\\
&&= \f{M}{2 \pi \hbar s} \; e^{- i \f{M}{2 \hbar s} R^2} \int \!\! dx \; e^{i \f{M}{\hbar s} Rx } \;V^{a_j}_{n_j 0} (x) \int \!\! dy \;  f(y) \; e^{i \f{M}{2 \hbar s} y^2 - i \f{M}{\hbar s}xy}
\nonumber\\
&&= \f{M}{2 \pi \hbar s} \; e^{- i \f{M}{2 \hbar s} R^2} \int \!\! dy \; f(y)
e^{i\f{M}{2 \hbar s} y^2} \int \!\! dx\; V^{a_j}_{n_j 0} (x) \;e^{-i \left( \f{My}{\hbar s} - \f{MR}{\hbar s}\right)x}\nonumber\\
&&= \f{M}{\sqrt{2 \pi} \hbar s} \; e^{- i \f{M}{2 \hbar s} R^2} \int \!\! dy \; f(y)
\;e^{i\f{M}{2 \hbar s} y^2}\;  \widetilde{V^{a_j}_{n_j 0}} \Big( \! M (\hbar s)^{-1}(y-R) \! \Big)\nonumber\\
&&= \f{1}{\sqrt{2 \pi} \gamma} \; \int \!\! d\xi \; f\! \big( R+ (M \gamma)^{-1} \hbar s \,\xi  \big) \; e^{i \f{R}{\gamma} \xi +i \f{\hbar s}{2 M \gamma^2}  \xi^2}  \;\widetilde{V^{a_j}_{n_j 0}} (\gamma^{-1} \xi)
\ea
where in the last line we have introduced the new integration variable $\xi =M \gamma(\hbar s)^{-1}(y-R)$.

\n
Using the convolution property of Fourier transform we have

\ba\label{Vn1}
&&V^{a_j}_{n_j 0} (x) = \f{1}{\gamma} \int \!\! dy \; \phi_{n_j} \big( \gamma^{-1}(y-a_j) \big) \phi_{0} \big( \gamma^{-1}(y-a_j) \big) V\big( \gamma^{-1} (x-y) \big)\nonumber\\
&&=  \int \!\! dz \; V(z)\; \big( \phi_{n_j}  \phi_0 \big)\! \big( \gamma^{-1} (x-a_j - \gamma z) \big)
\nonumber\\
&&= \gamma \int \!\! d k \;\tilde{V}(\gamma  k) \;\widetilde{(\phi_{n_j}  \phi_0)} (\gamma k) \; e^{i (x-a_j) k}
\ea
Hence
\be\label{trasV}
\widetilde{V^{a_j}_{n_j 0}}(k) = \sqrt{2 \pi} \gamma  \;\tilde{V}(\gamma  k) \;\widetilde{(\phi_{n_j}  \phi_0)} (\gamma k) \; e^{-i a_j k}
\ee

\n
Using (\ref{trasV}) in (\ref{freep}) and introducing the large parameter $\Lambda_j = \f{a_j}{\gamma}$ we conclude the proof. 

\vs
\hfill $\Box$

\vs
\vs

\n
Using the above lemma we can rewrite also  the integral in (\ref{hj}).

\vs
\n
{\bf Lemma 2.2.} {\em

\ba\label{iden}
&&
h_j^{\pm}(t) 
=\int_0^t \!\! ds \!\int \! \!d \xi \; F_j^{\pm}(\cdot, \xi,s) \;e^{i \Lambda_j \theta_j^{\pm}(\xi,s)} 
\ea
where 
\ba
&&F_j^{\pm}(R,\xi,s) =g_{j}(\xi) \, e^{i \f{\hbar s}{2 M \gamma^2} \xi^2}  \hat{\psi}_1^{\pm} (R, \xi,s) 
\label{F}\\
&&\hat{\psi}_1^{\pm} (R,\xi,s)= \f{\mathcal{N}}{\sqrt{\sigma}} e^{- \f{(R-\hat{R}_1)^2}{2 \sigma^2} \pm i \f{\hat{P}_1^{\pm}}{\hbar}R}
\label{psi1xis}\\
&&\hat{R}_1 =- \f{\hbar  }{M \gamma}\,  \xi s ,\;\;\;\;\;\;\hat{P}^{\pm}_1=P_0 \pm \f{\hbar }{\gamma}\xi 
\label{RP}\\
&& \theta_{j}^{\pm}(\xi,s)= \big( \!\pm \f{s}{\tau_j} -1 \big) \xi - q_j  \f{s}{\tau_j}  \label{tetajpm}
\ea
and $q_j$ has been defined in (\ref{qj}).}

\vs
\n
{\bf Proof.}  The proof is trivial if we notice that

\be\label{intra}
\psi^{\pm}\big( R+ (M\gamma)^{-1} \hbar s\, \xi  \big)  \; e^{i \f{R}{\gamma} \xi }= \hat{\psi}^{\pm}_1(R, \xi,s) \; e^{\pm i \Lambda_j \xi \f{s}{\tau_1}}
\ee

\n
and use lemma 2.1.

\vs
\hfill $\Box$

\vs
\vs

\n
The next step is to estimate   (\ref{iden}), i.e.  an integral containing the rapidly oscillating phase $\Lambda_j \theta^{\pm}_j (\xi, s)$.  
The   standard stationary (or non-stationary) phase methods can be used to obtain the estimate.  
 
\n
It is worth  mentioning that the integral in (\ref{iden}) contains also other phase factors depending on $(\xi,s)$ which, however, are slowly varying under our assumptions on the physical parameters of the model.

\n
The asymptotic  analysis for $\Lambda_j \rightarrow \infty$  is  simplified by the fact that  $\theta_{j}^{\pm}(\xi, s)$ is a quadratic function. 
The only critical points of the phase are  $(\pm q_j, \pm \tau_j)$ and, moreover, the hessian matrix 
is non degenerate, with eigenvalues $\pm \tau_j^{-1}$. 
This means that the behaviour of (\ref{iden}) for $\Lambda_j \rightarrow \infty$ in the case with  $\theta_j^-$ is radically different from the case with $\theta_j^+$, due to the fact that in the first case the critical point never belongs to the domain of integration while in the second case this happens for $t>\tau_j$.

\n
For the analysis of this second case it will be useful the  following  elementary lemma. For the convenience of the reader a proof of the lemma  will be given in the appendix.

\vs
\n
{\bf Lemma 2.3.} {\em Let us consider for any $\Lambda>0$

\be\label{J}
\mathcal{J}(\Lambda)= \int \! dx \int_{- \nu}^{\mu} \!\! dy \; f(x,y) \; e^{i \Lambda xy}
\ee

\n
where $ \mu, \nu$ are positive parameters, $f$ is a complex-valued, sufficiently smooth function. 
Then

\ba\label{J1}
\mathcal{J}(\Lambda)&= &  \f{1}{\Lambda} \, \mathcal{K}_1 (\Lambda)\\
&=&\f{1}{\Lambda} \, 2\pi f(0,0)+ \f{1}{\Lambda^2} \, \mathcal{K}_2 (\Lambda) \label{J2}\\
&=& \f{1}{\Lambda} \, 2\pi f(0,0)+ \f{1}{\Lambda^2} \, 2 \pi i \, d_{x} d_y f(0,0)+  \f{1}{\Lambda^3} \, \mathcal{K}_3 (\Lambda) \label{J3}
\ea

\n

\n
where $\mathcal{K}_l(\Lambda) $, $l=1,2,3$,  are explicitely given (see the appendix) and satisfy the estimates

\ba\label{stk1}
&&|\mathcal{K}_1 (\Lambda)| \leq c_1 \left( \|f(\cdot,\!0)\|_{L^1}
   + \int \!\! dx \|d_x d_y f(x,\!\cdot)\|_{L^2} \! \right)
\ea

\ba\label{stk2}
&&|\mathcal{K}_2 (\Lambda)| \leq  c_2 
 \left(      \|d_x^2 f(\cdot, \!0) \|_{L^1} + \| d_x d_y f(\cdot, \!0) \|_{L^1} + \int \!\! dx \, \| d_x^2 d_y^2 f(x, \!\cdot) \|_{L^2}
 \right)
\ea

\ba\label{stk3}
&&|\mathcal{K}_3 (\Lambda)| \! \leq \! c_3 \!\left( \! \|d_x^3 f(\cdot, \!0) \|_{L^1} \!+\! \| d_x^3 d_y f(\cdot, \!0) \|_{L^1} \!+\! \| d_x^2 d_y^2 f(\cdot, \!0) \|_{L^1} \! + \! \! \int \!\! \!dx  \| d_x^3 d_y^3 f(x, \!\cdot) \|_{L^2} \!\!
 \right)
\ea

\n
and the constants $c_1, c_2, c_3$ depend only on $\mu, \nu$}.

\vs
\vs
\n
Exploiting lemma 2.3 we obtain the following  asymptotic behaviour of (\ref{iden}) for $t>\tau_j$ when the phase is $\theta_j^+$.

\vs
\n
{\bf Proposition 2.4.} {\em For any  $t>\tau_j$ we have

\ba\label{F+}
&&h^+_j (t)= 
\f{2 \pi \tau_j}{\Lambda_j} \,  e^{-i \Lambda_j q_j} F_j^+ \!(\cdot, q_j, \tau_j)+ \f{1}{\Lambda_j^2} \mathcal{R}_j^{+} \!(\cdot,t, \Lambda_j)
\ea

\n
where

\ba\label{stR+j}
&&|\mathcal{R}^{+}_{j}(R,t,\Lambda_j)| \leq \! C_j  \bigg[ 
\int \!\! d \xi \, |d_{\xi}^{2} F^+_j  (R, \xi,\tau_j)| + \int \!\! d \xi \, |d_{\xi} d_s F^+_j (R,\xi,\tau_j)|    \nonumber\\
&& + \int \!\! d \xi \left( \int_0^t \!\!\!\! ds \, |d_{\xi}^2 d_s^2 F^+_j (R, \xi,s)|^2 \right)^{1/2} \bigg]
\ea
and $C_j$ depends on $t$ and $\tau_j$.}

\vs
\n
{\bf Proof.}  Let us introduce the change of coordinates $x=\xi - q_j$, $y = s - \tau_j$ in (\ref{iden})  and the shorthand notation $F(x,y)= e^{-i \Lambda_j q_j} F_j^+ (R, x+q_j, y+ \tau_j)$.  Then

\be\label{Fxy}
h^{+}_{j}(t)=  \int \!\!dx \int_{-\tau_j}^{t-\tau_j} \!\! dy \; F (x,y) \;e^{i\f{\Lambda_j}{\tau_j}xy} 
\ee

\n
The integral in (\ref{Fxy}) has the same form as the integral  (\ref{J}) analysed in  lemma 2.3,   if we identify  $\nu, \mu, f, \Lambda$ with $\tau_j, t-\tau_j, F, \f{\Lambda_j}{\tau_j}$ respectively.  Then, exploiting formula (\ref{J2}), we obtain the r.h.s. of (\ref{F+}) with 

\ba\label{R+j}
&&\mathcal{R}^+_j (R,t,\Lambda_j) = - \tau_j^2 \int\!\! dx \, \f{ F(x,0) - F(0,0) - d_x F(x,0)x}{x^2} \left(  \f{  e^{i \f{\Lambda_j}{\tau_j} (t-\tau_j) x}}{t- \tau_j}  + \f{  e^{- i \Lambda_j  x}}{ \tau_j}\right)\nonumber\\
&&+ \tau_j^2 \!\! \int \!\! dx \, d_x d_y F(x,\!0) \f{ e^{i \f{\Lambda_j}{\tau_j} (t- \tau_j)x}- e^{-i \Lambda_j x}}{x} \nonumber\\
&&-  \tau_j^2 \!\!\int_{- \tau_j}^{t-\tau_j} \!\!\!\!\!\! dy \!\int \!\! dx \f{ d_{x}^2 F(x,y) - d_{x}^2  F(x,0) - d_{x}^2 d_y F(x,0) y}{y^2} \, e^{i \f{\Lambda_j}{\tau_j}xy}
\ea
 
\n
Using (\ref{stk2}) we immediately get the estimate (\ref{stR+j}) and this concludes the proof.

\vs
\hfill$\Box$

\vs
\vs

\n
In the next  proposition we shall analyze the asymptotic behaviour of (\ref{iden}) when the phase is $\theta_j^-$. Taking into account the error term in (\ref{F+}),    it is sufficient to show  that $h_j^- (t) = O(\Lambda_j^{-2})$; on the other hand we remark that, following the same line, it is easy to extend the result to  $h_j^- (t) = O(\Lambda_j^{-k})$, for any integer $k$. 

\vs
\n
{\bf Proposition 2.5 .} {\em For any $t>0$ we have

\ba\label{F-}
&&
h_j^-(t) = \f{1}{\Lambda_j^2} \, \mathcal{R}^{-}_{j}(\cdot ,t,\Lambda_j)
\ea
where

\ba\label{Rj-}
&&
|\mathcal{R}_j^- (R,t,\Lambda_j) | \leq \int_0^t \!\!\! ds \! \int \!\! d \xi \left| d^2_{\xi} 
F_j^- (R,\xi,s) \right|
\ea

\n


}
\vs
\n
{\bf Proof.} If we notice that

\be
e^{i \Lambda_j \theta_j^- (\xi,s)} = \f{1}{ \left[-i \Lambda_j \left(\f{s}{\tau_j} +1 \right)\right]^2} \, d^2_{\xi}
e^{i \Lambda_j \theta_j^- (\xi,s)}
\ee

\n
and  integrate by parts two times in the r.h.s. of (\ref{iden}) we easily obtain the r.h.s. of (\ref{F-}) with

\ba\label{Rj-2}
&&\mathcal{R}^{-}_{j}(R,t,\Lambda_j)= - \tau_j^2 \int_0^t \!\! ds \f{1}{(s+\tau_j)^2} \int \! d \xi \left( d^2_{\xi}
F_j^- (R,\xi,s)  \right)  e^{i \Lambda_j \theta^{-}_{j}(\xi,s)}
\ea

\n
Then by a trivial estimate  we conclude the proof.

\vs
\hfill$\Box$

\vs
\vs
\n
Collecting together the results of propositions 2.4 and 2.5 we finally obtain an asymptotic expression for $t> \tau_j$   of the first order terms when $\Lambda_j \rightarrow \infty$

\ba\label{f1nj}
&&f^{(1)}_{n_j 0}(t) = \f{\mathcal{A}_{j}^{(1)}}{\Lambda_j} \; e^{- \f{i}{\hbar} t K_0}  \psi^{+}_j  + \f{1}{\Lambda_j^2} \mathcal{R}_{j}^{(1)}(\cdot,t , \Lambda_j)\\
&&\mathcal{A}_{j}^{(1)}= -  2 \pi i \,    \f{\lambda \tau_j}{\hbar}\; e^{-i (n_j +1) \omega t - i \Lambda_j q_j + i \f{\hbar \tau_j}{2 M \gamma^2}q_j^2} \;
g_{j}(q_j) \\
&&\psi_j^+ = \hat{\psi}_{1}^{+}(\cdot, q_j, \tau_j)\\
&&\mathcal{R}_{j}^{(1)}(\cdot,t , \Lambda_j)= - \Gamma_{n_j 0} (t) \left(\mathcal{R}_j^-  (\cdot,t, \Lambda_j) + \mathcal{R}^+_j (\cdot, t, \Lambda_j) \right)
\ea

\n
We observe that the leading term in the r.h.s. of (\ref{f1nj}) can  be more conveniently  written in the form

\ba\label{lt1}
&&\f{\mathcal{A}_{j}^{(1)}}{\Lambda_j} \; e^{- \f{i}{\hbar} t K_0}  \psi^{+}_j = - 2 \pi i \f{\lambda_0}{\sqrt{\delta m \, \delta E}} e^{i \eta_j(t)} \tilde{V}(q_j) \widetilde{\phi_{n_j} \phi_0}(q_j) e^{- \f{i}{\hbar} tK_0}  \psi_j^+\\
&&\eta_j(t)= \f{n_j^2}{2} \f{\delta E}{\delta \tau_j} -  (n_j + 1) \omega t +  \f{n_j}{\delta \tau_j}\\
&&\psi^+_j (R)= \f{b}{\sqrt{\sigma}} \, e^{- \f{(R-R_j)^2}{2 \sigma^2} + i \f{P_j}{\hbar}R}, \;\;\;\;\; R_j= n_j a_j \, \delta E  ,\;\;\;\; P_j = P_0 (1- n_j \, \delta  E)
\ea

\n
Then it is clear that the leading term has the form of a free evolution of  a wave packet which starts at $t= \tau_j$ from the position $a_j$ of $j^{th}$ oscillator, with   mean momentum $P_j $. Notice that under our assumptions $P_j  \simeq P_0 >0$. 
 
\n
In particular  (\ref{f1nj}) gives a precise meaning to the qualitative statement concerning the approximate behavior of $f^{(1)}_{n_j 0}(t)$ made in section 1.


\vs

\section{Second order approximation }

\vs
\n
In this section we fix $t> \tau_2$ and consider  the second order terms  corresponding to both oscillators in some exited states, i.e. terms of the type (see formula (\ref{2f2}))

\ba\label{hjl}
&&   - \int_0^t \!\! ds \; \Gamma_{n_j n_l} (t-s) \; V^{a_l}_{n_l 0} \; f^{(1), \pm}_{n_j 0} (s) \nonumber\\
&&=i\f{\lambda}{\hbar} \; \Gamma_{n_l n_j}(t)   \int_0^t \!\! ds \, e^{i n_l \omega s} \int_0^s \!\! ds' \, e^{i n_j \omega s'} \mathcal{I}_l (s) \mathcal{I}_j (s') \psi^{\pm}\\
&&\equiv i\f{\lambda}{\hbar} \; \Gamma_{n_l n_j}(t) \;h_{jl}^{\pm}(t)
\label{hjl1}
\ea

\n
for $j,l=1,2, \;\; l \neq j $. Proceeding as in lemmas 2.1 and 2.2, a straightforward computation in the case $a_2 >0$ yields

\ba\label{hjl2}
&&h_{jl}^{\pm}(t) = \int_0^t \!\! ds  \int_0^s \!\! ds'  \int \!\! d \xi \int \!\!d \eta \,
\, G_{jl}^{\pm} ( \cdot, \xi,s',\eta,s) \, 
e^{i \Lambda_j \theta_j^{\pm} (\xi,s')  + i \Lambda_l \theta_l^{\pm} (\eta,s)}\\
&&G_{jl}^{\pm} ( R, \xi,s',\eta,s) = g_{j}(\xi) g_{l}(\eta)  e^{i \f{\hbar }{2M \gamma^2} ( s' \xi^2+ s \eta^2 +  2 s \xi \eta)} \, \hat{\psi}_2^{\pm} (R, \xi,s',\eta,s)\\
&&\hat{\psi}_2^{\pm} (R,\xi,s',\eta,s)= \f{\mathcal{N}}{\sqrt{\sigma}} e^{-\f{(R-\hat{R}_2)^2}{2 \sigma^2} \pm i \f{\hat{P}_2^{\pm}}{\hbar} R}\\
&&\hat{R}_2 = - \f{\hbar}{M \gamma} (\xi s'+ \eta s) , \;\;\;\;\;\; \hat{P}_2^{\pm}= P_0  \pm \f{\hbar}{\gamma} (\xi + \eta)
\ea

\n
where $g_{j}$ and $\theta^{\pm}_j$ have been defined in (\ref{gnj}) and (\ref{tetajpm}) respectively. 
In the case $a_2 <0$ the same representation formula (\ref{hjl2}) holds if we replace $\Lambda_2$, $\tau_2$ with $- \Lambda_2$, $- \tau_2$, where $\Lambda_2=|a_2| \gamma^{-1}$, $\tau_2 = |a_2| v_0^{-1}$. 
In both cases, we shall  discuss the asymptotic behaviour of $h^{\pm}_{jl}(t)$ for $\Lambda_1, \; \Lambda_2 \rightarrow \infty$.

\n
The integral (\ref{hjl2}) contains  a rapidly oscillating phase and moreover the phase has exactly one critical point. Therefore  the behaviour strongly depends on whether or not the critical point lies in the integration domain. We shall  analyze separately the two cases $a_2 >0$ and $a_2 <0$.

\vs
\vs

3.1.   {\bf The case $a_2 >0$.} 

\vs
\n
We distinguish the four possible cases: i) $h_{21}^+(t)$,  ii) $h_{21}^-(t)$,  iii) $h_{12}^-(t)$,   iv) $h_{12}^+(t)$.  It is easily seen that the  point $(\xi_0, s'_0, \eta_0, s_0)$ where  the phase is stationary is: $(q_2,\tau_2, q_1, \tau_1)$ for  i), 
 $(-q_2, -\tau_2, -q_1, -\tau_1)$ for ii)  and $(-q_1, -\tau_1, -q_2, -\tau_2)$ for iii). 
 In all three  cases the stationary  point of the phase  does not belong to the domain of integration and then the integral rapidly decreases to zero for $\Lambda_1,  \Lambda_2 \rightarrow \infty$.  On the other hand in the case iv) the stationary point is $(q_1, \tau_1, q_2, \tau_2)$, i.e. it belongs to    the domain of integration  and therefore there is a leading term of order $(\Lambda_1 \Lambda_2)^{-1}$ which we shall  compute. 
In the next proposition we study the cases iv) 
following the same line of the proof of proposition 2.4.

\vs
\n
{\bf Proposition 3.1}. {\em For $a_2 >0$ and  $t>\tau_2$ we have
\ba\label{h12+}
&&h_{12}^+ (t) = \f{4 \pi^2 \tau_1 \tau_2}{\Lambda_1 \Lambda_2} e^{-i \Lambda_1 q_1 -i \Lambda_2 q_2} G_{12}^+ (\cdot, q_1, \tau_1, q_2, \tau_2)  + \f{1}{\Lambda_1^3} \, \mathcal{R}_{12}^{+} (\cdot,t,\Lambda_1, \Lambda_2)
\ea
where $ \mathcal{R}_{12}^{+} (R,t,\Lambda_1, \Lambda_2) $ is a bounded function of $\Lambda_1, \Lambda_2$, whose estimate will be given during the proof.

}

\vs
\n
{\bf Proof}.  Let us introduce the change of coordinates $x=\xi -q_1$, $y=s'-\tau_1$, $w=\eta -q_2$, $z=s-\tau_2$ in  (\ref{hjl2})  and the shorthand notation $G(x,y,w,z)=e^{-i\Lambda_1 q_1 -i \Lambda_2 q_2}G_{12}^+(R, x+q_1, y+\tau_1, w+q_2, z+\tau_2)$. Then

\ba\label{h12+1}
&&h_{12}^+(t) = \int_{-\tau_2}^{t-\tau_2} \!\! \!\!\!dz \! \int_{-\tau_1}^{z+\tau_2-\tau_1} \!\! \!\!\!\!dy  \! \int\!\!dx \! \int\!\!dw \,  G(x,y,w,z)e^{i \f{\Lambda_1}{\tau_1} xy +i \f{\Lambda_2}{\tau_2} wz}\\
&&=G(0,0,0,0) \lim_{a,b \rightarrow \infty}  \int_{-\tau_2}^{t-\tau_2} \!\! \!\!\!dz \! \int_{-\tau_1}^{z+\tau_2-\tau_1} \!\! \!\!\!\!dy  \! \int_{-a}^{a}\!\!\!dx \! \int_{-b}^{b}\!\!\!dw \,  e^{i \f{\Lambda_1}{\tau_1} xy +i \f{\Lambda_2}{\tau_2} wz} \nonumber\\
&&+  \lim_{a,b \rightarrow \infty}  \int_{-\tau_2}^{t-\tau_2} \!\! \!\!\!dz \! \int_{-\tau_1}^{z+\tau_2-\tau_1} \!\! \!\!\!\!dy  \! \int_{-a}^{a}\!\!\!dx \! \int_{-b}^{b}\!\!\!dw  \left( G(x,y,0,0) - G(0,0,0,0) \right)  e^{i \f{\Lambda_1}{\tau_1} xy +i \f{\Lambda_2}{\tau_2} wz} \nonumber\\
&&+ \lim_{a,b \rightarrow \infty}  \int_{-\tau_2}^{t-\tau_2} \!\! \!\!\!dz \! \int_{-\tau_1}^{z+\tau_2-\tau_1} \!\! \!\!\!\!dy  \! \int_{-a}^{a}\!\!\!dx \! \int_{-b}^{b}\!\!\!dw  \left( G(x,y,w,z) - G(x,y,0,0) \right)  e^{i \f{\Lambda_1}{\tau_1} xy +i \f{\Lambda_2}{\tau_2} wz} \nonumber\\
&&\equiv (I) + (II)+ (III)
\ea

\n
The term $(I)$ is the leading term and it can be easily computed

\ba\label{I}
&&(I)= \f{4 \pi^2 \tau_1 \tau_2}{\Lambda_1 \Lambda_2} G(0,0,0,0)  \lim_{a,b \rightarrow \infty}  \int_{-\tau_2}^{t-\tau_2} \!\! \!\!\!dz \! \int_{-\tau_1}^{z+\tau_2-\tau_1} \!\! \!\!\!\!dy \; \f{\sin \f{\Lambda_2}{\tau_2}bz}{\pi z} \; 
\f{ \sin \f{\Lambda_1}{\tau_1}ay}{\pi y}\nonumber\\
&&=\f{4 \pi^2 \tau_1 \tau_2}{\Lambda_1 \Lambda_2} G(0,0,0,0) \nonumber\\
&&=\f{4 \pi^2 \tau_1 \tau_2}{\Lambda_1 \Lambda_2} e^{-i \Lambda_1 q_1 -i \Lambda_2 q_2} G^{+}_{12} (R, q_1, \tau_1, q_2, \tau_2)
\ea

\n
Concerning the term $(II)$ we have 

\ba\label{II}
&&(II)= \f{2 \pi \tau_2}{\Lambda_2} \lim_{a,b \rightarrow \infty} \int_{-\tau_2}^{t-\tau_2}\!\!\! \!\!dz \; \f{\sin \f{\Lambda_2}{\tau_2} bz}{\pi z} \! \int_{-\tau_1}^{z+\tau_2-\tau_1} \!\! \!\!\!\!dy  \! \int_{-a}^{a}\!\!\!dx \! \left( G(x,y,0,0) - G(0,0,0,0) \right)  e^{i \f{\Lambda_1}{\tau_1} xy}\nonumber\\
&&= \f{2 \pi \tau_2}{\Lambda_2} \int\!\! dx 
 \! \int_{-\tau_1}^{\tau_2-\tau_1} \!\! \!\!\!\!dy   \left( G(x,y,0,0) - G(0,0,0,0) \right)  e^{i \f{\Lambda_1}{\tau_1} xy}
\ea

\n
The  r.h.s. of  (\ref{II})  can be estimated using  (\ref{J2}), (\ref{stk2}) and the result is 

\ba\label{II1}
&&|(II)| \leq  \f{ C_2 }{\Lambda_2 \Lambda_1^2} \bigg( \! \| d_x^2 G(\!\cdot ,\!0,\!0,0)\|_{L^1} + \| d_x d_y G(\!\cdot ,\!0,\!0 ,\!0)\|_{L^1} + \int \!\! dx \| d_x^2 d_y^2 G(\!x,\! \cdot ,\! 0,\!0) \|_{L^2}\! \bigg)\nonumber\\
&&= \f{ C_2 }{\Lambda_2 \Lambda_1^2}\bigg[ \int \!\! d \xi \, |d_{\xi}^2 G^{+}_{12} (R, \xi, \tau_1,q_2,\tau_2) | + \int \!\! d \xi \, | d_{\xi} d_{s'} G^{+}_{12} (R, \xi, \tau_1, q_2,\tau_2)|\nonumber\\
&&+ \int \!\! d \xi \, \left( \int_0^t \!\!\!\! ds' \, |d_{\xi}^2 d_{s'}^2 G^{+}_{12} (R,\xi,s',q_2,\tau_2)|^2 \right)^{1/2} \bigg]
\ea

\n
where $C_2$ is a constant depending on $\tau_1$, $\tau_2$. 
The term $(III)$ can be more conveniently written as

\ba\label{III}
&&(III)= \int \!\! dw \int_{- \tau_2}^{t- \tau_2} \!\!\!\!\!\! dz \, L(w,z) e^{i \f{\Lambda_2}{\tau_2} wz}\\
&&L(w,z)\equiv \int \!\!dx \int_{-\tau_1}^{z+\tau_2 - \tau_1} \!\!\!\!\!\! dy \, \big( G(x,y,w,z)- G(x,y,0,0) \big) \, e^{i \f{\Lambda_1}{\tau_1}xy}
\label{L}
\ea

\n
where $L(0,0)=0$. Using  (\ref{J3}), (\ref{stk3}) we have

\ba\label{III1}
&&(III)= (IV) + (V) 
\ea
where
\ba\label{q}
&&(IV)=  \f{2 \pi i \tau_2^2}{\Lambda_2^2}\,  d_{w} d_z L (0,0)\\
&&|(V)| \leq  \f{ C_3}{\Lambda_2^3} \bigg(\! \| d_w^3 L(\cdot,\!0)\|_{L^1} \!+\! \| d_w^3 d_z L(\cdot,\!0)\|_{L^1} \!+\! \|d_w^2 d_z^2 L(\cdot,\!0)\|_{L^1} \!+\!\! \int \!\! dw \|d_w^3 d_z^2 L(w, \cdot)\|_{L^2} \!\!\bigg)\nonumber\\
&&
\ea

\n
and $C_3$ depends on $t$, $\tau_1$, $\tau_2$. 
Taking into account (\ref{L}) we also obtain

\ba\label{q1}
&&|(V)| \leq \f{C_3}{\Lambda_2^3}  \bigg(\! \|d_w^3 G(\cdot,\! \cdot,\!\cdot,\!0)\|_{L^1} \!+\! \|d_w^3 d_z G(\cdot,\! \cdot,\!\cdot,\!0)\|_{L^1} \!
+\! 
\|d_w^2 d_z^2 G(\cdot,\! \cdot,\!\cdot,\!0)\|_{L^1} \! \nonumber\\
&&+\! \! \int \!\!\!dw \!\! \int \!\!\! dx \!\! \int_{-\tau_2}^{t-\tau_2} \!\!\!\!\! dy \, \|d_w^3 d_z^3 G(x,\!y,\!w,\! \cdot) \|_{L^2} \!\!\bigg)\nonumber\\
&&= \f{C_3}{\Lambda_2^3} \bigg[ 
\int_{0}^{\tau_2} \!\!\!\! ds' \!\! \! \int \!\! d\xi \!\!
\int \!\! d \eta \, | d_{\eta}^{3} G^{+}_{12} (R,\xi,s',\eta,\tau_2)|
+ \int_{0}^{\tau_2} \!\!\!\! ds' \!\! \! \int \!\! d\xi \!\!
\int \!\! d \eta \, | d_{\eta}^{3}d_s G^{+}_{12} (R,\xi,s',\eta,\tau_2)| \nonumber\\
&&+ \int_{0}^{\tau_2} \!\!\!\! ds' \!\! \! \int \!\! d\xi \!\!
\int \!\! d \eta \, | d_{\eta}^{2}d^2_s G^{+}_{12} (R,\xi,s',\eta,\tau_2)| + \! \int_{0}^{t} \!\!\!\! ds' \!\! \! \int \!\! d\xi \!\!
\int \!\! d \eta \! \left(\! \int_0^t \!\!\!\! d s \, |d^3_{\eta} d^3_s G^+_{12} (R, \xi,s',\eta,s)|^2 \right)^{1/2} \bigg] \nonumber\\
&&
\ea

\n
Concerning $d_w d_z L(0,\!0)$,  a straightforward computation gives

\ba\label{III2}
&&d_{w} d_z L(0,0)= \int \!\! dx \, d_{w}G(x, \!\tau_2 - \tau_1,\!0,\!0) \, e^{i \f{\Lambda_1}{\tau_1}(\tau_2 - \tau_1)x} + \! \int\!\! dx \!\! \int_{-\tau_1}^{\tau_2-\tau_1} \!\!\!\!\!\! dy \,d_{w} d_z G(x,\!y,\!0,\!0) \, e^{i \f{\Lambda_1}{\tau_1} xy}\nonumber\\
&&\equiv (IV_1) + (IV_2)
\ea

\n
In (\ref{III2}) we integrate  by parts in the first integral  and use  (\ref{J1}), (\ref{stk1}) in the second integral. Then
\ba\label{III_1}
&&|(IV_1)| \leq \f{1}{\Lambda_1} \, \f{\tau_1}{\tau_2 - \tau_1} \| d_x d_w G(\cdot, \! \tau_2 - \tau_1, \!0, \!0) \|_{L^1} \nonumber\\
&&= \f{1}{\Lambda_1} \, \f{\tau_1}{\tau_2 - \tau_1} \! \int \!\! d \xi \, |d_{\xi} d_{\eta} G^+_{12} (\xi, \tau_2, q_2, \tau_2)|\\
&&|(IV_2)| \leq \f{C_1}{\Lambda_1} \bigg( \| d_w d_z G(\cdot, \!0,\!0, \!0) \|_{L^1} \!+ \! \int \!\!dx \| d_x d_y d_w d_z G(x, \!\cdot, \!0,\!0) \|_{L^2} \!\! \bigg)\nonumber\\
&&=\f{C_1}{\Lambda_1} \bigg[ \! \int \!\! d \xi \, |d_{\eta} d_s G^+_{12} ( \xi, \tau_1, q_2, \tau_2)| + \!  \int \!\! d \xi \! \left( \int_{0}^{\tau_2} \!\!\!\! d s' \, | d_{\xi} d_{s'} d_{\eta} d_{s}G^+_{12} (\xi,s',q_2,\tau_2)|^2 \! \right)^{\! 1/2} \! \bigg] 
\label{III_2}
\ea

\n
and $C_1$ depends on $\tau_1$, $\tau_2$. 
Taking into account (\ref{II1}), (\ref{q1}), (\ref{III_1}), (\ref{III_2}) we   get (\ref{h12+}), with an explicit estimate of $\mathcal{R}^{+}_{12}(R,t,\Lambda_1, \Lambda_2)$, and this  concludes the proof.

\vs
\hfill $\Box$

\vs
\vs
\n
Let us consider the cases i),ii),iii) where the stationary point of the phase lies out of the integration region. In such cases, exploiting repeated integration by parts, one can show that (\ref{hjl2}) is $O(\Lambda_1^k)$, for any integer $k$, for $\Lambda_1 \rightarrow \infty$. Since the error term in (\ref{h12+}) is $O(\Lambda_1^{-3})$, in the next proposition we shall limit to $k=3$.

\vs
\n
{\bf Proposition 3.2}.  {\em For $a_2 >0$,   $t>\tau_2$   we have
\ba\label{h21+}
&&h_{jl}^{a}(t) = \f{1}{\Lambda_1^3} \, \mathcal{R}_{jl}^{a}(\cdot,t,\Lambda_1, \Lambda_2) 
 \ea
for $a=\pm, j=2,l=1$ and $a=-,j=1,l=2$, 
where
\ba\label{R21+1}
&&|\mathcal{R}_{jl}^{a}(R,\! t,\!\Lambda_1,\! \Lambda_2\!)| \!  \leq   C
\!\! \int_0^t \!\!\!\! ds \!\! \int_0^s \!\!\!\! ds' \!\! \int \!\! \!d \xi \! \!
 \int \!\! \!d \eta \, \bigg( \left| d^3_{\eta} 
  G_{jl}^{a} (R,\xi,s'\!, \!\eta,\!s) \!  + \!  d_{\xi}^3 G_{jl}^{a} (R,\xi,s'\!,\! \eta,\! s) \right|  \bigg)  
\ea
and $C$ depends on $\tau_1, \tau_2$.

}

\vs
\n
{\bf Proof.}  Let us define $\tau_0=\f{\tau_1 + \tau_2}{2}$ and write

\ba\label{h21++}
&&h_{21}^+ (t)= \int_0^{\tau_0} \!\!\! ds \! \int_0^s \!\!\! ds' \! \! \int \!\! d \xi \! \int \!\! d \eta \, G_{21}^+ (\cdot, \xi,s', \eta,s) e^{i \Lambda_2 \theta_2^+ (\xi, s') + i \Lambda_1 \theta_1^+ (\eta,s)} \nonumber\\
&&+  \int_{\tau_0}^{t} \!\!\! ds \! \int_0^s \!\!\! ds' \! \! \int \!\! d \xi \! \int \!\! d \eta \, G_{21}^+ (\cdot, \xi,s', \eta,s) e^{i \Lambda_2 \theta_2^+ (\xi, s') + i \Lambda_1 \theta_1^+ (\eta,s)}\nonumber\\
&&\equiv (A) + (B)
\ea

\n
In $(A)$ we integrate by parts three times with respect to the variable $\xi$ and we obtain

\ba\label{h21+1}
&&\left| (A)\right|  = \left|  \f{\tau_2^3}{i^3 \Lambda_2^3} \int_0^{\tau_0}\!\!\!\! ds \!\! \int_0^s \!\!\!\! ds' \! \f{1}{(\tau_2 - s')^3} \! \int \!\! d \xi \! \int \!\! d \eta \left( d^3_{\xi} G_{21}^+ (\cdot, \xi,s',\eta,s) \right) e^{i \Lambda_2 \theta_2^+ (\xi, s') + i \Lambda_1 \theta_1^+ (\eta,s)}\right|  \nonumber\\
&&\leq  \f{1}{\Lambda_2^3} \, \f{\tau_2^3}{(\tau_2 - \tau_0)^3} \int_0^{\tau_0} \!\!\!\! ds \! \int_0^s \!\!\!\! ds' \! \int\!\! d \xi \!\int\!\! d \eta \left| d_{\xi}^3 G_{21}^+ (\cdot,\xi,s', \eta, s) \right|
\ea

\n
Integrating by parts with respect to $\eta$ in $(B)$ we have

\ba\label{h21+2}
&&|(B)|= \left|\f{(- \tau_1)^3 }{i^3 \Lambda_1^3}  \!\!\int_{\tau_0}^{t}\!\!\!ds \f{1}{(s-\tau_1)^3}  \! \int_0^s \!\!\! ds' \!  \int \!\! d \xi \! \int \!\! d \eta \left( d^3_{\eta} G_{21}^+ (\cdot, \xi,s',\eta,s) \right) e^{i \Lambda_2 \theta_2^+ (\xi, s') + i \Lambda_1 \theta_1^+ (\eta,s)}\right| \nonumber\\
&&\leq \f{1}{\Lambda_1^3} \, \f{\tau_1^3}{(\tau_0 - \tau_1)^3} \int_{\tau_0}^{t} \!\!\! ds \! \int_0^s \!\!\! ds' \! \! \int \!\! d \xi \! \int \!\! d \eta \, |d_{\eta}^3  G_{21}^+ (\cdot, \xi,s', \eta,s) |
\ea

\n
From (\ref{h21+1}) and (\ref{h21+2}) we get the estimate for $h_{21}^+(t)$.

\n
For the estimate of $h_{21}^- (t)$ it is sufficient to notice that 
\ba\label{ipk}
&&e^{i \Lambda_2 \theta_2^- (\xi,s') + i \Lambda_1 \theta_1^- (\eta,s)}= \f{1}{\left[ -i \Lambda_1 \left( \f{s}{\tau_1} +1 \right) \right]^3} \,  e^{i \Lambda_2 \theta_2^- (\xi,s')} d^3_{\eta} \, e^{  i \Lambda_1 \theta_1^- (\eta,s)}
\ea
and to integrate by parts three times. The estimate of $h_{12}^- (t)$ is analogous and then the proof is complete.

\vs
\hfill $\Box$

\vs
\vs
 
\vs
\n
Taking into account   (\ref{hjl1}), (\ref{h12+}), (\ref{h21+}), 
 we  obtain for $n_1, n_2 \neq 0$, $a_2 >0$ and $t > \tau_2$

\ba\label{f2fi}
&&f^{(2)}_{n_1 n_2} (t)= \f{\mathcal{A}^{(2)} }{\Lambda_1 \Lambda_2} e^{-\f{i}{\hbar} t K_0} \, \psi^{+}_{12} + \f{1}{\Lambda_1^3}\,   \mathcal{R}^{(2)} (\cdot, t, \Lambda_1 , \Lambda_2)\\ 
&&\mathcal{A}^{(2)}= - 4 \pi^2 \f{\lambda^2 \tau_1 \tau_2}{\hbar^2}  e^{ -i  (n_1 + n_2 +1) \omega t -i \Lambda_1 q_1 - i \Lambda_2 q_2} \, e^{i \f{\hbar }{2M \gamma^2} ( \tau_1 q_1^2 + \tau_2 q_2^2 +2 \tau_2 q_1 q_2)} \,g_{1}(q_1) g_{2}(q_2)\\
&&\psi^{+}_{12} = \hat{\psi}^{+}_2 (\cdot, q_1, \tau_1, q_2,\tau_2) \\
&&\mathcal{R}^{(2)} (\cdot, t, \Lambda_1 , \Lambda_2)=
i \f{\lambda}{\hbar} \Gamma_{n_1 n_2} (t) \!\! \!\sum_{\substack{a=\pm \\  j,l=1,2, j \neq l}}  \!\! \!\mathcal{R}^{a}_{jl} ( \cdot, t, \Lambda_1, \Lambda_2)\label{R2}
\ea

\n
Notice that the leading term in (\ref{f2fi}) can also be written as

\ba\label{lt2}
&& \f{\mathcal{A}^{(2)} }{\Lambda_1 \Lambda_2} e^{-\f{i}{\hbar} t K_0} \, \psi^{+}_{12}= 
- 4 \pi^2  \f{\lambda_0^2}{\delta m \, \delta E} \, e^{i \eta_{12}(t)} \prod_{j=1,2} \! \tilde{V}(q_j) \widetilde{(\phi_{n_j}  \phi_0)}(q_j) \,      e^{-\f{i}{\hbar} t K_0} \psi_{12}^+\\
&&\eta_{12}(t)= \f{n_1^2}{2} \f{\delta E}{\delta \tau_1} + \f{n_2^2}{2} \f{\delta E}{\delta \tau_2} + n_1 n_2 \f{\delta E}{\delta \tau_2} - (n_1 + n_2 +1) \omega t + \f{n_1}{\delta \tau_1} +  \f{n_2}{\delta \tau_2}\\
&&\psi_{12}^+ (R)= \f{\mathcal{N}}{\sqrt{\sigma}}\,  e^{- \f{(R-R_{12})^2}{2 \sigma^2} + i \f{P_{12}}{\hbar} R}, \;\;\;\;\; 
R_{12}= (n_1 a_1 \!+\! n_2 a_2 ) \, \delta E , \;\;\;\;  P_{12}= P_0 [1 - (n_1  + n_2) \, \delta E ] \label{psi12+}
\nonumber\\
&&
\ea

\vs
\vs

3.2.  {\bf The case $a_2 <0$.} 

\vs
\n
Here the two oscillators are on the opposite sides with respect to the origin and  one can easily check that the point $(\xi_0, s'_0, \eta_0, s_0)$ where the phase in (\ref{hjl2}) is stationary is: $(q_1, \tau_1, q_2, -\tau_2)$ for $h_{12}^{+}(t)$, 
$(-q_1, -\tau_1, -q_2, \tau_2)$ for $h_{12}^{-}(t)$, 
$(q_2, -\tau_2, q_1, \tau_1)$ for $h_{21}^{+}(t)$, 
$(-q_2, \tau_2, -q_1, -\tau_1)$ for $h_{21}^{-}(t)$. Since none of these points belongs to the domain of integration we can show that  $h_{jl}^{\pm}(t)$ is always rapidly decreasing to zero for $\Lambda_1, \Lambda_2 \rightarrow \infty$.

\vs
\n
{\bf Proposition 3.3}. {\em For $a_2 <0$, $t>\tau_2$ and any integer $k>2$ we have

\ba\label{2h12+}
&&h_{jl}^{\pm}(t)= \f{1}{\Lambda^k_1} \mathcal{Q}_{jl}^{\pm} (\cdot,t,\Lambda_1,\Lambda_2), \;\;\;\;\; j,l=1,2
\ea
where
\ba\label{Q12+}
&&|\mathcal{Q}_{jl}^{\pm} (R,t,\Lambda_1,\Lambda_2)| \leq \int_0^t \!\!\! ds \!\! \int_0^s \!\!\! ds' \!\!\int \!\!d \xi \!\!\int \!\! d \eta \, \bigg( |d_{\eta}^k G_{jl}^{\pm} (R, \xi,s',\eta,s)|  +  |d_{\xi}^k G_{jl}^{\pm} (R, \xi,s',\eta,s)| \bigg)\nonumber
    \\
&&
\ea

 }

\vs
\n
{\bf Proof.} The proof is an immediate consequence of $k$ integration by parts and a trivial estimate.

\vs
\hfill $\Box$

\vs
\vs
\n
From the above proposition we conclude that for $n_1, n_2 \neq 0$, $a_2 <0$, $t> \tau_2$ and any integer $k>2$ we have 

\ba\label{f2a2<}
&&f^{(2)}_{n_1 n_2} (t) = \f{1}{\Lambda_1^k} \mathcal{Q}^{(2)} (\cdot,t,\Lambda_1,\Lambda_2)\\
&& \mathcal{Q}^{(2)} (\cdot,t,\Lambda_1,\Lambda_2)= i \f{\lambda}{\hbar} \Gamma_{n_1 n_2}(t)
\!\!\! \sum_{\substack{a=\pm \\ j,l=1,2, j \neq l}}
 \!\!\!  \mathcal{Q}_{jl}^{a} (\cdot,t,\Lambda_1,\Lambda_2) \label{Q2}
 \ea


\vs

\section{Joint excitation probability}

\n
We are now in position to compute the joint excitation probability of the two oscillators in the two cases $a_2 <0$ and $a_2 >0$. As a preliminary step, we need  a pointwise  estimate of the derivatives of $G^{\pm}_{jl}$ with respect to the variables $\xi, \eta$. 

\n
It is convenient to introduce the following notation 

\ba\label{nota}
&&\mathfrak{a}= \f{\hbar t}{M \gamma^2}, \;\;\;\;  \;\;\mathfrak{b} = \f{\hbar t}{M \gamma \sigma}, \;\;\;\;  \mathfrak{c} = \f{\sigma}{\gamma}\\
&&s=t \alpha, \;\;\;\; \;\;\;\;\;\;s'=t \beta, \;\;\;\; \;\;\;\;\;\\
&&x =\sigma^{-1} R, \;\;\;\;\; z = x + \mathfrak{b}( \beta \xi + \alpha \eta) \label{xz}
\ea

\n
We notice that, for $t$ of the same order of magnitude of $\tau_2$, the constants in (\ref{nota}) are of order one; moreover the rescaled variables $\alpha, \beta$ satisfy $0 \leq \alpha, \beta \leq 1$.

\vs
\n
{\bf Lemma 4.1}.  {\em We have

\ba\label{deG}
&&  |d_{\eta}^k G_{jl}^{\pm} (R, \xi,t \beta,\eta,t \alpha)|  +  |d_{\xi}^k G_{jl}^{\pm} (R, \xi,t \beta,\eta,t \alpha)| \nonumber\\
&&\leq c \;\mathfrak{A}_k (t)\,  \f{\mathcal{N}}{\sqrt{\sigma}} \, \langle z \rangle^k \, e^{- \f{z^2}{2}} \,  \langle \xi  \rangle^{k} \!\sum_{m=0}^{k} | d^m_{\xi} g_j (\xi)|   \, \langle \eta \rangle^{k}\! \sum_{m=0}^{k} | d^m_{\eta} g_l (\eta)|  
\ea
where
\ba
&&\mathfrak{A}_k (t) = \left( 1 + \mathfrak{a}^2 + \mathfrak{b}^4 \right)^{k/2} \left( 1 + \mathfrak{b}^2 + \mathfrak{c}^2 \right)^{k/2} \left( 1+ \mathfrak{a} 
\right)^k
\ea

}

\vs
\n
{\bf Proof.} Exploiting the above notation we can write 
\ba\label{G2}
&&G^{\pm}_{jl} = \psi^{\pm} g_j \, g_l \,  e^{\phi}
\ea
where $\psi^{\pm}=\psi^{\pm}(R)$, $g_j=g_j(\xi)$, $g_l=g_l(\eta)$ and
\ba\label{fi}
&&\phi= i \mathfrak{a} \! \left(\! \f{\beta}{2} \xi^2 + \f{\alpha}{2} \eta^2 + \alpha \xi \eta \!\right) \!- \!\mathfrak{b}^2 \!\left( \!\f{\beta^2}{2} \xi^2 + \f{\alpha^2}{2} \eta^2 + \alpha \beta \xi \eta \! \right) \!- \!\mathfrak{b} x \!\left(\beta \xi + \alpha \eta \right) \!+\! i \mathfrak{c}x (\xi + \eta)
\ea

\n
Let us compute the derivative of order $k$ with respect to $\eta$.

\ba\label{deG2}
&&d^k_{\eta} G^{\pm}_{jl}=\psi^{\pm} g_j \sum_{m=0}^{k} 
\left( \!\!\!\begin{array}{c} 
m \\ k
\end{array} \!\!\! \right)
d_{\eta}^{k-m} g_l d_{\eta}^{m} e^{\phi} \nonumber\\
&&=\psi^{\pm} e^{\phi} g_j \sum_{m=0}^{k} 
\left( \!\!\!\begin{array}{c} 
m \\ k
\end{array} \!\!\! \right)
d_{\eta}^{k-m} g_l\!\!\sum_{\substack{n,p \\ n+2p=m}}\!
 \f{m!}{n!\,  p! \, 2^p}(d_{\eta}\phi)^n (d_{\eta}^2 \phi)^p
\ea
 
\n
A straightforward computation yields
\ba\label{dfi}
&&|d_{\eta} \phi |^n \leq \left[  \sqrt{\mathfrak{b}^2 + \mathfrak{c}^2} |z| + (\mathfrak{a}+ \mathfrak{b} \mathfrak{c}) (|\xi| + |\eta|) \right]^n \nonumber\\
&&= \sum_{q=0}^n \left( \!\!\!\begin{array}{c} 
n \\ q
\end{array} \!\!\! \right)
(\mathfrak{b}^2 + \mathfrak{c}^2 )^{\f{n-q}{2}} |z|^{n-q} (2 \mathfrak{a} 
)^q (|\xi|+|\eta|)^q \nonumber\\
&&\leq c \, (1 + \mathfrak{b}^2 + \mathfrak{c}^2 )^{k/2}  (1+ \mathfrak{a} 
)^k \langle z \rangle^k \langle \xi \rangle^k \langle \eta \rangle^k
\ea

\ba\label{d2fi}
&&|d_{\eta}^2 \phi|^p \leq (1+ \mathfrak{a}^2 + \mathfrak{b}^4)^{k/2}
\ea

\ba\label{psifi}
&&\left| \psi^{\pm} e^{\phi} \right| = \f{\mathcal{N}}{\sqrt{\sigma}} \, e^{- \f{z^2}{2}}
\ea

\n
Using (\ref{dfi}), (\ref{d2fi}), (\ref{psifi}) in (\ref{deG2}) we obtain the estimate
\ba\label{deG3}
&&|d^k_{\eta} G^{\pm}_{jl} | \leq c \, \mathfrak{A}_k (t) \f{\mathcal{N}}{\sqrt{\sigma}} \, \langle z \rangle^k \, e^{-\f{z^2}{2}}  \langle \xi \rangle^k |g_j(\xi)| \, \langle \eta \rangle^k \! \sum_{m=0}^k |d_{\eta}^{m} g_l (\eta) |
\ea

\n
Following exactly the same line we also find the corresponding  estimate of $|d_{\xi}^k G^{\pm}_{jl}|$ and this concludes the proof of the lemma.

\vs
\hfill $\Box$

\vs
\vs
\n
Finally we can prove our main result.

\vs
\n
{\bf Proof of theorem 1}. We start with a detailed estimate of $\mathcal{P}_{n_1 n_2}^{-}(t)$. Taking into account (\ref{f2a2<}), (\ref{Q2}), (\ref{Q12+}) we have

\ba\label{P-1}
&&\mathcal{P}_{n_1 n_2}^{-}(t) 
\leq \f{1}{\Lambda_1^{2k}} \int \!\! dR \, |\mathcal{Q}^{(2)} (R,t,\Lambda_1,\Lambda_2)|^2 \nonumber\\
&&\leq \f{4 \lambda^4}{\hbar^4 \Lambda_1^{2k}} 
\sum_{ \substack{a=\pm \\ j,l=1,2, j \neq l}} \int \!\! dR \, |\mathcal{Q}^{a}_{jl} (R,t, \Lambda_1,\Lambda_2)|^2 \nonumber\\
&&\leq \f{16 \lambda^4}{\hbar^4 \Lambda_1^{2k}} 
\sup_{a,j,l} \int \!\! dR \left[ \int_0^t \!\!\!\! ds \! \int_0^s \!\!\!\! ds'\! \int \!\! d \xi \! \int \!\! d \eta \left( | d_{\eta}^{k} G^a_{jl}| + |d^k_{\xi} G^a_{jl}| \right) \right]^2 \nonumber\\
&&\leq \f{16 \lambda^4}{\hbar^4 \Lambda_1^{2k}} 
\sup_{a,j,l} \left\{ \int_0^t \!\!\!\! ds \! \int_0^s \!\!\!\! ds'\! \int \!\! d \xi \! \int \!\! d \eta \left[ \int \!\! dR \, \left(      | d_{\eta}^{k} G^a_{jl}| + |d^k_{\xi} G^a_{jl}| \right)^2 \right]^{1/2} \right\}^2
\ea
where in the last line we have interchanged the order of integration and used the Schwartz inequality. Exploiting the estimate (\ref{deG}) we find 
\ba\label{P-2}
&&\mathcal{P}^{-}_{n_1 n_2}(t) \leq \f{c}{\Lambda_1^{2k -4}} \,\f{\lambda^4}{\hbar^4 \Lambda_1^4}\, \mathfrak{A}^2_k (t) \mathcal{N}^2  \|g_1\|^2_{W^{k,1}_k} \|g_2\|^2_{W^{k,1}_k} \! \left[   \int_0^t \!\!\!\! ds \! \int_0^s \!\!\!\! ds'\! \left( \f{1}{\sigma} \int \!\! dR \, \langle z \rangle^{2k} e^{-z^2} \right)^{1/2} \right]^2 \nonumber\\
&&\leq \f{c}{\Lambda_1^{2k -4}} \,\f{\lambda^4 t^4}{\hbar^4 \Lambda_1^4}\, \mathfrak{A}^2_k (t) \mathcal{N}^2  \|g_1\|^2_{W^{k,1}_k} \|g_2\|^2_{W^{k,1}_k} \nonumber\\
&&= \f{c}{\Lambda_1^{2k -4}} \left(\! \f{\lambda_0}{\sqrt{\delta m  \delta E}} \! \right)^{\!\!4} \! \left( \! \f{t}{\tau_2} \! \right)^{\!\!4} \! \left(\! \f{a_2}{a_1} \! \right)^{\!\!4} \! \mathfrak{A}^2_k (t) \, \mathcal{N}^2  \|g_1\|^2_{W^{k,1}_k} \|g_2\|^2_{W^{k,1}_k}
\ea

\n
It remains to evaluate the two norms in (\ref{P-2}). Recalling the definition of $g_j(\xi)$ (see (\ref{gnj})) we have
\ba\label{deg}
&&\|g_j\|_{W^{k,1}_k} \leq \sum_{m=0}^k \sum_{p=0}^{m}
\left( \!\!\! \begin{array}{c}  m \\ p  \end{array} \!\!\! \right)  \int \!\! d \xi \, \langle \xi \rangle^k \left| 
d^{m-p}_{\xi} \tilde{V}(\xi) \, d^p_{\xi} (\widetilde{\phi_{n_j} \phi_0 )} (\xi) \right| \nonumber\\
&&\leq \sum_{m=0}^k \sum_{p=0}^{m}
\left( \!\!\! \begin{array}{c}  m \\ p  \end{array} \!\!\! \right)  
\f{1}{\sqrt{2 \pi}} \int \!\! dx |x|^{m-p} |V(x)| \int \!\! d \xi \, \langle \xi \rangle^k \left| 
 d^p_{\xi} (\widetilde{\phi_{n_j} \phi_0 )} (\xi) \right| \nonumber\\
 &&\leq c \, \|V\|_{L^1_k} \| \widetilde{\phi_{n_j} \phi_0 }\|_{W_k^{k,1}}
\ea


\n
Inserting (\ref{deg}) in (\ref{P-2}) we finally get the estimate (\ref{p-}) with
\ba\label{P-3}
&&C_{n_1 n_2}^{(k)}(t) \equiv c \left(\! \f{\varepsilon}{\sqrt{\delta m \delta E}} \! \right)^{\!4} \left( \! \f{t}{\tau_2} \! \right)^{\!\!4} \! \left(\! \f{a_2}{a_1} \! \right)^{\!\!4} \! \mathfrak{A}^2_k (t) \, \mathcal{N}^2 \|V\|_{L^1_k}^4 \| \widetilde{\phi_{n_1} \phi_0 }\|_{W_k^{k,1}}^2  \| \widetilde{\phi_{n_2} \phi_0 }\|_{W_k^{k,1}}^2
\ea

\n
Let us consider $\mathcal{P}^+_{n_1 n_2}(t)$. From (\ref{f2fi}), (\ref{R2}), (\ref{lt2}), (\ref{psi12+}) we have

\ba\label{P+}
&&\mathcal{P}^+_{n_1 n_2}(t) = 16 \pi^4  \sqrt{\pi} \left( \!\! \f{\lambda_0}{\sqrt{\delta m \delta E}} \!\!\right)^{\!\!4} \mathcal{N}^2  \,|g_1(q_1) g_2(q_2)|^2 + \mathcal{S}_{n_1 n_2}(t)
\ea
 
\n
where $\mathcal{S}_{n_1 n_2}(t)$ is a correction term of order $\Lambda_1^{-1}$. In fact

\ba\label{S}
&&|\mathcal{S}_{n_1 n_2}(t)| \leq \f{c}{\Lambda_1^3}  \, \f{\lambda_0^2}{\delta m \delta E} \,\mathcal{N}\,  |g_1(q_1) g_2(q_2)|
 \left(  \int \!\! dR \,  |\mathcal{R}^{(2)}(R,t,\Lambda_1,\Lambda_2)|^2  \right)^{\!1/2}
\nonumber\\
&&+ \f{1}{\Lambda_1^6} \int \!\! dR \,  |\mathcal{R}^{(2)}(R,t,\Lambda_1,\Lambda_2)|^2 \nonumber\\
&&\leq \f{c}{\Lambda_1} \, \f{\lambda_0^2}{\delta m \delta E}  \,\f{\lambda^2 }{\hbar^2 \Lambda_1^2}\,\mathcal{N}\,  |g_1(q_1) g_2(q_2)|  \left(  \sup_{a,j,l} 
  \int \!\! dR \,  |\mathcal{R}^{a}_{jl}(R,t,\Lambda_1,\Lambda_2)|^2  \right)^{\!1/2}
\nonumber\\
&&+ \f{c}{\Lambda_1^2}\,  \f{\lambda^4}{\hbar^4 \Lambda_1^4} \sup_{a,j,l} \int \!\! dR \,  |\mathcal{R}^{a}_{jl}(R,t,\Lambda_1,\Lambda_2)|^2 \nonumber\\
&&= \f{1}{\Lambda_1} \left( \! \f{\lambda_0}{\varepsilon} \! \right)^{\!\!4} D_{n_1 n_2} (t)
\ea
where
\ba\label{Dt}
&&D_{n_1 n_2} (t) \equiv c \left(  \! \f{\varepsilon}{ \sqrt{ \delta m  \delta E}} \! \right)^{\! \!4}   \left( \! \f{t}{\tau_2} \! \right)^{\!\!2} 
\!\! \left( \! \f{a_2}{a_1} \! \right)^{\!\!2} \! \bigg[ \mathcal{N}\,  |g_1(q_1) g_2(q_2)|  \left(  \f{1}{t^4} \sup_{a,j,l} 
  \int \!\! dR \,  |\mathcal{R}^{a}_{jl}(R,t,\Lambda_1,\Lambda_2)|^2  \right)^{\!\!1/2} \nonumber\\
&&+ \f{1}{\Lambda_1} \left( \! \f{t}{\tau_2} \! \right)^{\!\!2} 
\!\! \left( \! \f{a_2}{a_1}\! \right)^{\!\!2}    \f{1}{t^4} \sup_{a,j,l} 
  \int \!\! dR \,  |\mathcal{R}^{a}_{jl}(R,t,\Lambda_1,\Lambda_2)|^2  
  \bigg]
\ea

\n
The proof of (\ref{p+}), (\ref{p++}) is complete if we notice that  the quantity

\ba\label{S1}
&& \f{1}{t^4 } \sup_{a,j,l} \int \!\! dR \,  |\mathcal{R}^{a}_{jl}(R,t,\Lambda_1,\Lambda_2)|^2
\ea

\n
can be estimated following the line of the previous case. The explicit computation is straightforward but   rather long and tedious and  we omit the details.

\vs
\hfill $\Box$


\section{Appendix}

\vs
\n
Here we give a proof of lemma 2.3 (see e.g. \cite{bh} for analogous computations).

\vs

\n
{\bf Proof of lemma 2.3.} Let  us decompose $\mathcal{J}(\Lambda)$ in the following form

\ba
&&\mathcal{J}(\Lambda)= \int\!\! dx \int_{- \nu}^{\mu} \!\!\!\! dy f(x,0) \,e^{i \Lambda xy} + \int\!\! dx \int_{- \nu}^{\mu} \!\!\!\! dy \, (f(x,y)-f(x,0)) e^{i \Lambda xy}\nonumber\\
&&=- \f{i}{\Lambda} \int\!\! dx \, f(x,0) \,  \f{e^{i \Lambda \mu x} - e^{-i\Lambda \nu x}}{x} + \f{i}{\Lambda} \int \!\! dx \! \int_{-\nu}^{\mu} \!\!\!\!\!d y \,  \f{d_x f(x,y) - d_x f(x,0)}{y} \, e^{i \Lambda xy}\nonumber\\
&&\equiv \f{1}{\Lambda} \left( \mathcal{K}_{11}(\Lambda) + \mathcal{K}_{12}(\Lambda) \right)
\ea

\n
where an explicit integration in the first integral and an integration by parts in the second integral has been performed. Thus we have (\ref{J1}) with $\mathcal{K}_{1}(\Lambda) =  \sum_{j=1}^{2}\mathcal{K}_{1j}(\Lambda)$.  The estimate of $\mathcal{K}_{1}(\Lambda)$ is easily obtained if we write 
\be 
 d_x f(x,y) - d_x f(x,0) \!=  y \!\int_{0}^{1} \!d\theta \, d_x d_y  f(x,y \theta) 
\ee

\n
 and then use the Schwartz inequality.

\n
In order to prove (\ref{J2}) we reconsider $\mathcal{K}_{11}(\Lambda)$ and $\mathcal{K}_{12}(\Lambda)$. In particular we have 

\ba\label{K11}
&&\mathcal{K}_{11}(\Lambda)= - i f(0,0) \int \!\!dx \, \f{e^{i \Lambda \mu x}-e^{-i\Lambda \nu x}}{x}   -i \int\!\! dx \left(f(x,0)- f(0,0)\right) \f{e^{i \Lambda \mu x}-e^{-i\Lambda \nu x}}{x} \nonumber\\
&&=2 \pi f(0,0) - \f{1}{\Lambda} \int \!\! dx \, \f{f(x,0) - f(0,0) - d_x f(x,0) x}{x^2} \left( \f{e^{i \Lambda \mu x}}{\mu} + \f{e^{-i \Lambda \nu x}}{\nu} \right)\nonumber \\
&&\equiv 2 \pi f(0,0) + \f{1}{\Lambda} \, \mathcal{K}_{21}(\Lambda)
\ea

\n
where we have explicitely computed the first integral and we have integrated by parts in the second integral. Concerning $\mathcal{K}_{12}(\Lambda)$, we observe that it is of the same form as $\mathcal{J}(\Lambda)$ and then we can repeat the procedure. Denoting $\eta (x,y) \equiv \f{d_x f(x,y) - d_x f(x,0)}{y} \, e^{i \Lambda xy}$, with $\eta(x,0)=d_x d_y f(x,0)$, we obtain

\ba\label{K12}
&&\mathcal{K}_{12}(\Lambda)= \f{1}{\Lambda} \int\!\! dx \, \eta(x,0) \,  \f{e^{i \Lambda \mu x} - e^{-i\Lambda \nu x}}{x} -  \f{1}{\Lambda} \int \!\! dx \! \int_{-\nu}^{\mu} \!\!\!\!\!d y \,  \f{d_x \eta(x,y) - d_x  \eta(x,0)}{y} \, e^{i \Lambda xy}\nonumber\\
&&= \f{1}{\Lambda}\! \int\!\! dx \, d_x d_y f(x,\!0) \,  \f{e^{i \Lambda \mu x} - e^{-i\Lambda \nu x}}{x} -  \f{1}{\Lambda} \! \int \!\! dx \! \int_{-\nu}^{\mu} \!\!\!\!\!d y \,  \f{d_x^2  f(x,\!y) - d_x^2  f(x,\!0) - d_x^2 d_y f(x,\!0) y}{y^2} \, e^{i \Lambda xy}\nonumber\\
&&\equiv \f{1}{\Lambda} \left( \mathcal{K}_{22}(\Lambda) + \mathcal{K}_{23}(\Lambda) \right)
\ea

\n
and the asymptotic formula (\ref{J2})  follows, with $\mathcal{K}_2 (\Lambda) = \sum_{j=1}^3 \mathcal{K}_{2j}(\Lambda)$.

\n
The estimate of $\mathcal{K}_{21}(\Lambda)$ is obtained if we write 
\be
f(x,\!0)- f(0, \!0) - d_x f(x, \!0)x = -x^2 \! \int_0^1 \!d \theta \, \theta d_x^2 f(x \theta , \!0),
\ee

\n
 the estimate of $\mathcal{K}_{22}(\Lambda)$ is trivial and for  $\mathcal{K}_{23}(\Lambda)$ it is sufficient to notice that 
 \be
 d_x^2 f(x, \!y) - d_x^2 f(x, \!0) - d_x^2 d_y f(x, \!0) y = y^2 \!\int_0^1 \!\!d \theta \, \theta  \!\int_0^1 d \zeta \,  d_x^2 d_y^2 f(x, y \theta \zeta)
 \ee
 
 \n
 and to use the Schwartz inequality. Then the estimate 
(\ref{stk2}) for $\mathcal{K}_2 (\Lambda) $ is proved.

\n
Finally we shall prove (\ref{J3}). An integration by parts in $\mathcal{K}_{21}(\Lambda)$ yields

\ba\label{K21}
&&\mathcal{K}_{21}(\Lambda)=  \f{2i}{\Lambda}\! \int \!\! dx \, \f{f(x,\!0) - f(0,\!0) - d_x f(x,\!0) x + d_x^2 f(x,\!0) \f{x^2}{2}}{x^3} \!\left(\! \f{ e^{i \Lambda \mu x}}{\mu^2} - \f{ e^{-i \Lambda \nu x}}{\nu^2} \!\right)\nonumber\\
&&\equiv \f{1}{\Lambda} \, \mathcal{K}_{31}(\Lambda)
\ea

\n
For $\mathcal{K}_{22}(\Lambda) $ we proceed as in (\ref{K11}) and we obtain

\ba\label{K22}
&&\mathcal{K}_{22}(\Lambda)= d_x d_y f(0,\!0)\! \int \!\!dx \f{e^{i \Lambda \mu x} \!-\! e^{-i \Lambda \nu x}}{x} + \!\int \!\! dx \left( d_x d_y f(x, \!0) \!-\! d_x d_y f(0, \!0) \right)\! \f{e^{i \Lambda \mu x} \!-\! e^{-i \Lambda \nu x}}{x}\nonumber\\
&&= 2 \pi i d_x d_y f(0, \!0) - \f{i}{\Lambda} \int \!\! dx \f{d_x d_y f(x, \!0) - d_x d_y f(0, \!0) - d_x^2 d_y f(x, \!0) x}{x^2} \left( \f{ e^{i \Lambda \mu x}}{\mu} + \f{e^{-i \Lambda \nu x}}{\nu} \right)\nonumber\\
&&\equiv 2 \pi i d_x d_y f(0, \!0) + \f{1}{\Lambda} \,\mathcal{K}_{32}(\Lambda)
\ea

\n
The last term $\mathcal{K}_{23}(\Lambda)$ has the same form as $\mathcal{J}(\Lambda)$ and then following the same argument we get

\ba\label{K23}
&&\mathcal{K}_{23}(\Lambda)= \f{i}{2 \Lambda} \int \!\! dx \, d_x^2 d_y^2 f(x,\!0) \f{e^{i \Lambda \mu x} - e^{-i \Lambda \nu x}}{x} \nonumber\\
&&- \f{i}{\Lambda} \int \!\! dx \int_{-\nu}^{\mu} \!\!\!\!\! dy \f{ d_x^3 f(x, \!y) - d_x^3 f(x, \!0) - d_x^3 d_y f(x, \!0) y - d_x^3 d_y^2 f(x, \!0) \f{y^2}{2}}{y^3} \,e^{i \Lambda xy}\nonumber\\
&&\equiv \f{1}{\Lambda} \, ( \mathcal{K}_{33}(\Lambda) + \mathcal{K}_{34}(\Lambda)  )
\ea

\n
and (\ref{J3}) is proved with $\mathcal{K}_{3}(\Lambda) = \sum_{j=1}^4 \mathcal{K}_{3j}(\Lambda)$.

\n
The estimate (\ref{stk3}) for $\mathcal{K}_{3}(\Lambda) $ is easily obtained following the same line as before and we omit the details.

\vs

\hfill $\Box$

\vs

\vs
\vs
\vs


\begin{thebibliography}{99}
\vs


\bibitem[Be]{be} Bell J.S., Quantum mechanics for cosmologists. In {\em Speakable and unspeakable in quantum mechanics}, Cambridge University Press, Cambridge, 1987. 

\bibitem[BPT]{p} Blasi R., Pascazio S., Takagi S., Particle tracks and the mechanism of decoherence in a model bubble chamber. {\em Phys. Lett. A}, {\bf 250}, 230-240 (1998).


\bibitem[BH]{bh} Bleinstein N., Handelsman R.A., {\em Asymptotic Expansions of Integrals}, Dover Publ., New York, 1975.

\bibitem[Br]{b} Broyles A.A., Wave mechanics of particle detectors. {\em Phys. Rev. A}, {\bf 48}, n. 2, 1055-1065 (1993).

\bibitem[CCF]{ccf} Cacciapuoti C., Carlone R., Figari R.
A solvable model of a tracking chamber. 
{\em Rep. Math. Phys.}, {\bf 59} (2007).

\bibitem[CL]{cl} Castagnino M, Laura R., Functional approach to quantum decoherence and the classical final limit: the Mott and cosmological problems. {\em Int. J. Theo. Phys.}, {\bf 39}, n. 7, 1737-1765 (2000).

\bibitem[HA]{ha} Halliwell J.J., Trajectories for the wave function of the universe from a simple detector model. {\em Phys. Rev. D}, {\bf 64}, 044008 (2001).

\bibitem[H]{h} Heisenberg W., {\em The Physical Principles of Quantum Theory}, Dover Publ., New York, 1951.

\bibitem[LR]{lr} Leone M., Robotti N., A note on the Wilson cloud chamber (1912). {\em Eur. J. Phys.}, {\bf 25}, 781-791(2004). 

\bibitem[M]{m} Mott, N.F., The wave mechanics of $\alpha$-ray tracks. {\em Proc. R. Soc.
Lond.}, {\bf A 126}, 79-84 (1929).


\end{thebibliography}
\end{document}